\documentclass[10pt]{iopart}

\usepackage{iopams}
\usepackage{color}
\usepackage{graphicx} 
\usepackage{dcolumn}  
\usepackage{bm}       

\renewcommand{\vec}[1]{\mathbf{#1}}
\newcommand{\mean}[1]{\left\langle #1 \right\rangle}

\begin{document}

\title[Nonequilibrium properties of strongly correlated artificial atoms]{Nonequilibrium properties of strongly correlated artificial atoms---a Green's functions approach}

\author{K Balzer and M Bonitz}
\address{Institut f\"ur Theoretische Physik und Astrophysik, Christian-Albrechts-Universit\"at Kiel, Leibnizstrasse 15, 24098 Kiel, Germany}
\ead{balzer@theo-physik.uni-kiel.de}


\begin{abstract}
A nonequilibrium Green's functions (NEGF) approach for spatially inhomogeneous, strongly correlated artificial atoms is presented and applied to compute the time-dependent properties while starting from a (correlated) initial few-electron state at finite temperatures. In the regime of moderate to strong coupling, we consider the Kohn mode of a three-electron system in a parabolic confinement excited by a short pulsed classical laser field treated in dipole approximation. In particular, we numerically confirm that this mode is preserved within a conserving (e.g.~Hartree-Fock or second Born) theory .
\end{abstract}

\vspace{2pc}

\section{Introduction\label{sec1}}
\begin{figure}[t]
\begin{center}
\includegraphics[width=0.3\textwidth]{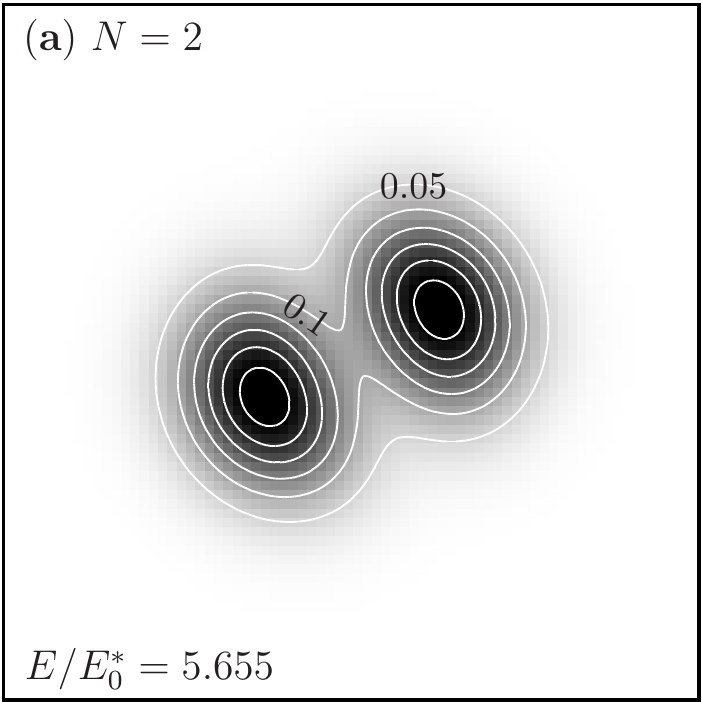}
\includegraphics[width=0.3\textwidth]{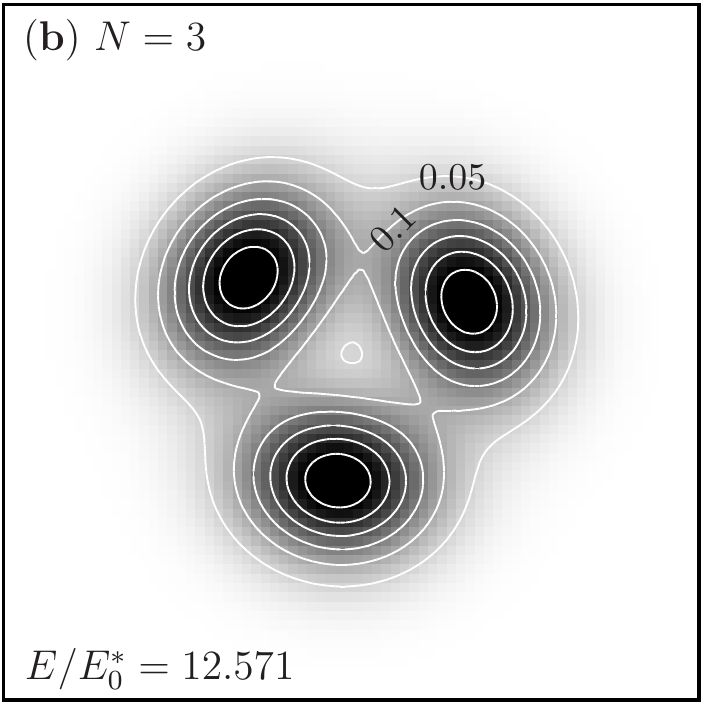}
\includegraphics[width=0.3\textwidth]{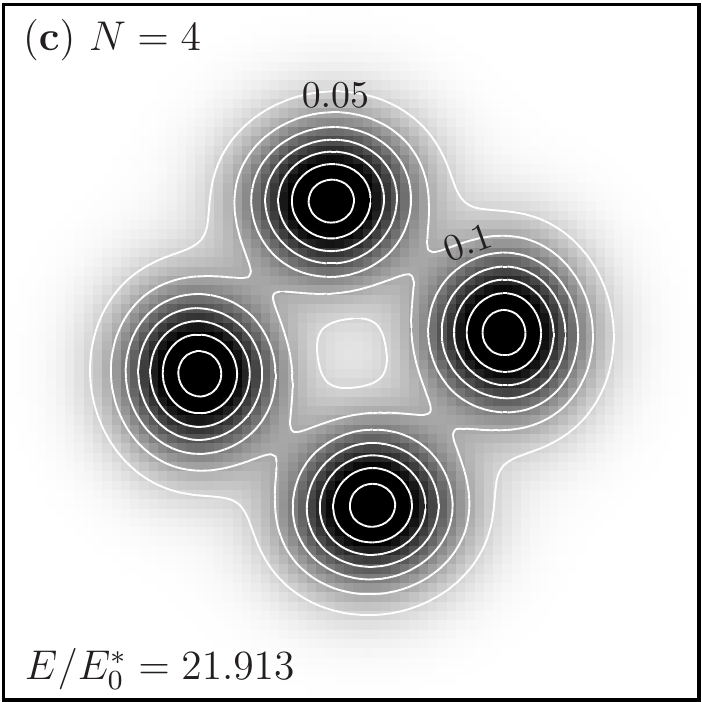}\\
\includegraphics[width=0.3\textwidth]{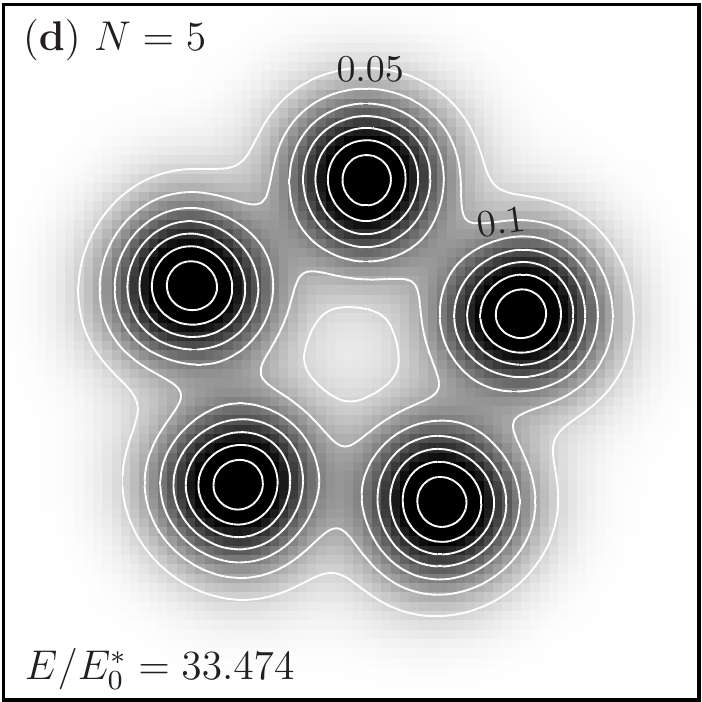}
\includegraphics[width=0.3\textwidth]{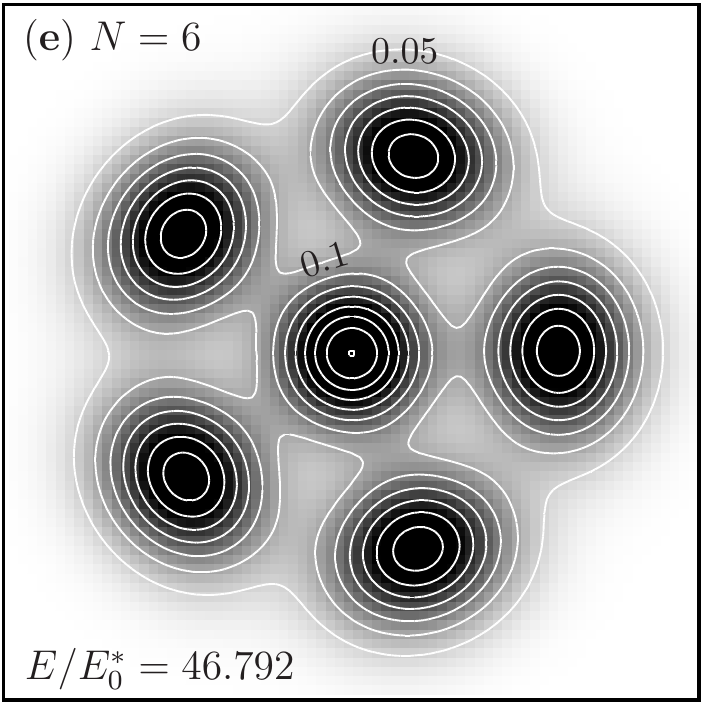}
\includegraphics[width=0.3\textwidth]{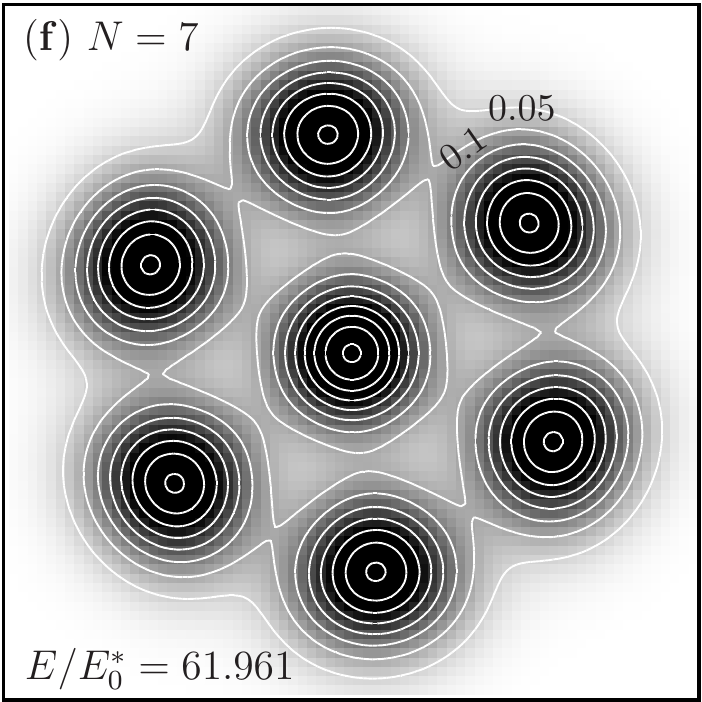}
\end{center}
\caption{Examples for the one-electron ground state (GS) density $n^\mathrm{GS}_N(\vec{r})$ of $N=2,3,\ldots,7$ electrons in an isotropic 2D trap for $\beta=100$ and $\lambda=5.0$. Figures (\textbf{a}) to (\textbf{f}) display the result of (symmetry broken) unrestricted Hartree-Fock calculations. The side length of the density plots measures $8x_0^*$.}\label{Fig1}
\end{figure}

'Artificial atoms' (AA) are inhomogeneous quantum few-particle systems confined in a trapping potential and show bound (discrete) electronic states, as they are occurring in real atoms\cite{ashoori96}. Most artificial atoms are realized in an (isotropic) parabolic confinement and quantum dots are a synonym convention for these systems, e.g.~Refs~\cite{jauregui93,balzer08}. But AAs are also formed by ions in Penning and Paul traps, charge carriers in semiconductor heterostructures (quantum wells), or electrons in metal clusters. AAs in one- (1D) and two-dimensional (2D) entrapment show interesting properties far from ideal Fermi-gas behavior including ring and shell structures---see the 2D ground state configurations for $N=2,3,\ldots, 7$ electrons displayed in Fig.~\ref{Fig1}. This is due to the externally controllable electron-electron interactions which, in particular, induce correlation phenomena.

Previous theoretical investigations of AAs mainly concern ground state calculations and approaches to thermodynamic equilibrium, see Refs.~\cite{balzer08,filinov01} and references. The aim of this paper is to present a time-dependent theory including moderate and strong correlations. Following a quantum statistical approach, we thereby want to consider the system's response to a (strong) short-pulsed laser field while being embedded into an environment of finite temperature. To this end, in Sec.~\ref{sec2} and \ref{sec3}, we describe the NEGF method applied to compute the (correlated) initial state which is then propagated in time according to the Keldysh/Kadanoff-Baym equations. In Sec.~\ref{sec4}, we present results for a three-electron AA in a 1D and 2D trap geometry.

\section{Model\label{sec1.1}}
The $d$-dimensional $N$-electron Hamiltonian of the artificial atom in a time-dependent laser field $\vec{E}(t)$ described in dipole approximation reads
\begin{eqnarray}\label{ham}
 \hat{H}(t)&=&\sum_{i=1}^{N}\left(-\frac{\hbar^2}{2 m_e^*}\nabla_{\!i}^2+\frac{m_e^*}{2} \omega_0^2 \vec{r}_i^2+e \vec{E}(t)\vec{r}_i\right)+\sum_{i<j}^{N}\frac{e^2}{4\pi\varepsilon\, r_{ij}}\;,\;\;\;
\end{eqnarray}
where the effective electron mass is given by $m_e^*$, the frequency $\omega_0$ adjusts the confinement strength (and hence the density in the AA), $e$ is the elementary charge and $\varepsilon$ denotes the background dielectric constant. Further, the $d$-dimensional electron coordinates $\vec{r}_i$ originate from the trap centre and $r_{ij}=|\vec{r}_i-\vec{r}_j|$. 


Hamiltonian (\ref{ham}) can be rewritten in dimensionless form, using the transformations $\{E\rightarrow E/E_0^*, \vec{r}\rightarrow\vec{r}/x_0^*\}$, where $E^*_0=\hbar\omega_0$ is the confinement energy and the (oscillator) length $x_0^*=\sqrt{\hbar/(m^*_e\omega_0)}$ denotes the characteristic spatial one-electron extension in the AA. Consequently, the system is characterized by a single coupling (or Wigner) parameter $\lambda$, which relates the characteristics Coulomb energy $E_C=e^2/(4\pi\varepsilon x_0^*)$ to $E^*_0$ according to
\begin{eqnarray}\label{lambda}
 \lambda=\frac{E_C}{E^*_0}=\frac{e^2}{4\pi\varepsilon\,x_0^* \hbar\omega_0}=\frac{x_0^*}{a_B}\;,
\end{eqnarray}
with $a_B$ being the effective electron Bohr radius. The Hamiltonian (\ref{ham}) then transforms into dimensionless form
\begin{eqnarray}\label{hamx0}
 \hat{H}_{\lambda}(t)&=&\frac{1}{2}\sum_{i=1}^{N}\left(-\nabla_{\!i}^2+\vec{r}_i^2+\gamma(t)\vec{r}_i\right)\,+\,\lambda\sum_{i<j}^{N}\frac{1}{r_{ij}}\;,
\end{eqnarray}
with $\gamma(t)=eE(t)/\sqrt{\hbar m^*_e\omega_0}$. The coupling parameter $\lambda$ adjusts the influence of electron-electron interactions and (quantum) correlations: In the case of $\lambda\ll1$, the artificial atom behaves similar to an ideal Fermi gas. For $\lambda\sim1$, the equilibrium state of the AA is Fermi liquid-like, whereas in the limit $\lambda\rightarrow\infty$, it is $x_0^*\gg a_B$, and quantum effects vanish in favor of classical interaction dominated charge carriers \cite{ludwig08}. For moderate coupling ($\lambda\gtrsim 1$) the AAs typically show spatially well localized carrier density including Wigner molecule (Wigner crystal-like) behavior\cite{filinov01}, see Fig.~\ref{Fig1}. Moreover, if the AA is not in its ground state (GS), one has to take into account thermodynamic fluctuations due to a surrounding heat bath of dimensionless temperature $\beta^{-1}=k_B T/E_0^*$. Below, all presented results are related to the system of units $\{x_0^*, E_0^*\}$ and in the definition of the NEGFs we take $\hbar=1$.


\section{Preparation of equilibrium states\label{sec2}}
Introducing electron annihilation (creation) operators $\psi^{(\dagger)}(\vec{r}_1t_1)$ acting in the Heisenberg picture at a space-time point $1=(\vec{r}_1,t_1)$, the second-quantized form of (\ref{hamx0}) reads
\begin{eqnarray}\label{hamsq}
H_\lambda(t_1)&=&\int \textup{d}^dr\,\hat{\psi}^\dagger(1)\,h^0(1)\,\hat{\psi}(1)\\
&&+\int\!\!\!\int\textup{d}^dr\,\textup{d}^d\bar{r}\,\hat{\psi}^\dagger(1)\,\hat{\psi}^\dagger(\bar{1})\,w(\vec{r}_1-\vec{r}_{\bar{1}})\,\hat{\psi}(\bar{1})\,\hat{\psi}(1)\;,\nonumber
\end{eqnarray}
with the one-electron energy $h^0(1)=(-\nabla^2_{r_{1}}+\vec{r}_1^2)/2+\gamma(t_1)\vec{r}_1$ and the interaction $w(\vec{r}_1-\vec{r}_{\bar{1}})=\lambda\,|\vec{r}_1-\vec{r}_{\bar{1}}|^{-1}$. In the following, we study Hamiltonian (\ref{hamsq}) at finite temperatures $\beta^{-1}$ by means of the one-particle nonequilibrium Green's function $G(1,\bar{1})$, which is defined on the Schwinger/Keldysh contour $\cal C$ (see e.g.~Refs.~\cite{qkt98,keldysh64}) as
\begin{eqnarray}
 G(1,\bar{1})=-i\langle T_{\cal C}\,\hat{\psi}(1)\,\hat{\psi}^\dagger(\bar{1})\rangle\;,
\end{eqnarray}
where $T_{\cal C}$ denotes time-ordering on $\cal C$. $G(1,\bar{1})$ obeys the two-time Keldysh/Kadanoff-Baym equation (KBE)\cite{kadanoff62}
\begin{eqnarray}
\label{KBE1}
 [\,i\partial_{t_1}\!-\!h(1)\,]\,G(1,\bar{1})=\delta_{\cal C}(1\!-\!\bar{1})-i\!\int_{\cal C}\!\!\textup{d}2\,W(1\!-\!2)G_{12}(1, 2;\bar{1},2^+)\;,
\end{eqnarray}
and its adjoint, where $W(1-\bar{1})=\delta_{\cal C}(t_1-t_{\bar{1}})w(\vec{r}_1-\vec{r}_{\bar{1}})$ acts instantaneously with contour delta function $\delta_{\cal C}$, and $2^+$ indicates the time limit $t_2\rightarrow t_2+0$. Here, the two-particle Green's function,
\begin{eqnarray}
 G_{12}(1,2;\bar{1},\bar{2})&=&(-i)^2\langle T_{\cal C}\,\hat{\psi}^\dagger(1)\,\hat{\psi}^\dagger(\bar{1})\,\hat{\psi}^\dagger(\bar{2})\,\hat{\psi}^\dagger(2)\rangle\;,
\end{eqnarray}
appears as a generalization of the two-particle density matrix. In terms of $G_{12}$, we can formulate all relevant many-body approximations: For instance, substituting $G_{12}(1,2;\bar{1},\bar{2})\rightarrow G(1,\bar{1})G(2,\bar{2})-G(1,\bar{2})G(2,\bar{1})$ yields the Hartree-Fock (HF) approximation. Second and higher order approximations (e.g.~second Born or GW), can be systematically obtained by diagram expansions known from ground state many-body theory and are valid for both equilibrium and nonequilibrium situations.

When in Eq.~(\ref{hamsq}), we consider $\gamma(t_1)\equiv0$ for $-\infty<t_1\leq t_0$, the AA stays in thermodynamic equilibrium until at a time $t_1>t_0$ $\gamma(t_1)$ becomes nonzero. Without loss of generality we thereby can take $t_0=0$. Specifying the time-independent single-electron part of Eq.~(\ref{hamsq}) as $h^0(\vec{r}_1)=h(1)|_{\gamma(t_1)=0}$, the KBE then reduce, for $t_{1,\bar{1}}\leq0$, to the Dyson equation
\begin{eqnarray}\label{deq}
 &&[\,-\partial_\tau-h^0(\vec{r}_1)\,]\,G^M\!(\vec{r}_1,\vec{r}_{\bar{1}};\tau)\\
&=&\delta(\tau)+\!\int\!\!\textup{d}^2\bar{r}\int_0^\beta\!\!\textup{d}\bar{\tau}\,\Sigma^M_\lambda(\vec{r}_1,\bar{\vec{r}};\tau-\bar{\tau})\,G^M\!(\bar{\vec{r}},\vec{r}_{\bar{1}};\bar{\tau})\;.\nonumber
\end{eqnarray}
Here, the Matsubara Green's function being defined as $G^M(1,\bar{1})=G^M(\vec{r}_1,\vec{r}_{\bar{1}};\tau)=G(\vec{r}_1 0-i\tau_1,\vec{r}_{\bar{1}} 0-i\tau_{\bar{1}})$, with $\tau=\tau_1-\tau_{\bar{1}}\in[-\beta,\beta]$, characterizes the equilibrium (initial) state of the AA. Further, on the right hand side, we have introduced the self-energy $\Sigma_\lambda^M(1,\bar{1})=\Sigma^M_\lambda(\vec{r}_1,\vec{r}_{\bar{1}};\tau)$ according to $-iW(1-2)G_{12}(1, 2;\bar{1},2^+)|_{t_{1,{\bar{1}}\leq0}}=\Sigma[G](1,2)G(2,\bar{1})$. A conserving many-body approximation\cite{kadanoff62}, i.e.~an approximation for $\Sigma$ that preserves density (continuity equation), total energy and momentum, can now be formulated in terms of a functional $\Phi$ such that $\Sigma(1,\bar{1})=\delta\Phi[G]/\delta G(\bar{1},1)$.

At a given temperature $\beta^{-1}$, most of the equilibrium properties of the AA system (\ref{ham}), e.g.~total energy, one-particle density and energy spectrum, are contained in $G^M$, see the formulas in Sec.~\ref{sec3.2} and take the limit $t_1\rightarrow0$. For the numerical techniques applicable in solving Eq.~(\ref{deq}) in matrix form see e.g. Refs.~\cite{balzer08,thesis07,dahlen05}. In HF approximation, the self-consistent solution can be written as
\begin{eqnarray}\label{HFgf}
 G^M(\vec{r}_1,\vec{r}_{\bar{1}};\tau)&=&\sum_{m=0}^{n_b-1}\phi_m^*(\vec{r}_1)\,\phi_m(\vec{r}_{\bar{1}})g^M_{mm}(\tau)\;,\\
 \hspace{2.35pc}g^M_{mm}(\tau)&=&f_\beta(\epsilon_m-\mu)\,e^{-\tau(\epsilon_m-\mu)}=e^{-\tau(\epsilon_m-\mu)}/(e^{\beta(\epsilon_m-\mu)}+1)\;,\nonumber
\end{eqnarray}
with interaction renormalized (effective single-electron) HF orbitals $\phi_m(\vec{r})$\cite{ludwig08}, quantum numbers $m=0,\ldots,n_b-1$, discrete energies $\epsilon_m$, and a chemical potential $\mu$. Beyond HF level, $G^M$ will be no longer diagonal in the functions $\phi_m$, and the respective occupation probabilities will deviate from a Fermi-Dirac distribution $f_\beta(\epsilon_m-\mu)$ due to additional electron scattering processes. In particular, the inclusion of electron-electron correlations leads to orbital-dependent energy shifts and broadening in the HF spectrum $a(\omega)=\sum_{m}\delta(\omega-\epsilon_m)$, see Ref.~\cite{balzer08}.


\section{Time-propagation of initial states\label{sec3}}
When for $t_1>0$ the laser field is switched on and $\gamma(t_1)\neq0$, the quantum state of the AA evolves in real time according to the KBE, Eq.~(\ref{KBE1}) and its adjoint. Thereby, being computed from the Dyson equation (\ref{deq}) in a self-consistent manner, the Matsubara Green's function serves as initial (Kubo-Martin-Schwinger) condition for the time-propagation. In particular, for $t_0=0$, one has
\begin{eqnarray}
G(\vec{r}_1 0-i\tau_1,\vec{r}_{\bar{1}} 0-i\tau_{\bar{1}})=i\,[\,G^M(\vec{r}_1,\vec{r}_{\bar{1}};\tau_1)- G^M(\vec{r}_1,\vec{r}_{\bar{1}};-\tau_{\bar{1}})]\;.
\end{eqnarray}
Beyond mean field level, all relevant initial correlations are taken into account via $G^M$ and, consequently, evolve in time, leading to a correlated $N$-particle dynamics.


\subsection{Solving the Keldysh-Kadanoff Baym equations\label{sec3.1}}
The expansion of the NEGF in terms of a HF basis, see Eq.~(\ref{HFgf}), advises us also to solve the real-time KBE in matrix form\cite{thesis07,dahlen06}. This means that we generally consider
\begin{eqnarray}\label{gfm} G(1,\bar{1})=\sum_{m,n=0}^{n_b-1}\phi^*_m(\vec{r}_1)\,\phi_n(\vec{r}_{\bar{1}})\,g_{mn}(t_{1},t_{\bar{1}})\;,
\end{eqnarray}
 with time arguments $t_1,t_{\bar{1}}$ on the contour ${\cal C}$, the coefficient matrix $g_{mn}(t_{1},t_{\bar{1}})=\theta(t_{1},t_{\bar{1}})g^>_{mn}(t_{1},t_{\bar{1}})-\theta(t_{\bar{1}},t_{1})g^<_{mn}(t_{1},t_{\bar{1}})$ of dimension $n_b\times n_b$, and steady-state HF orbitals $\phi_m(\vec{r})$ which generate a complete orthonormal set. Hence, $g_{mn}(t_1,t_{\bar{1}})=-i\langle T_{\cal C}\,c_m(t_1)\,c_n^\dagger(t_{\bar{1}})\rangle$ are just the NEGFs with respect to the operators $\hat{c}^\dagger_m$ ($\hat{c}_m$) that create (annihilate) an electron in the state $m$. Consequently, the diagonal elements [$m=n$] are directly related to the occupation numbers (probabilities) of the HF orbitals, cf.~Eq.~(\ref{occprob}), whereas the off-diagonal elements [$m\neq n$] are connected with 'interband' excitations, i.e.~the transition probabilities between the energy levels.

Inserting expression (\ref{gfm}), Eq.~(\ref{KBE1}) and its adjoint then transform into integro-differential equations for the matrix elements $g_{mn}(t_1,t_{\bar{1}})=(\vec{g})_{mn}(t_1,t_{\bar{1}})\,$: 
\begin{eqnarray}\label{KBE1mf}
\hspace{0.8pc}\left[\,i\partial_{t_{1}}-\vec{h}(t_1)\,\right]\vec{g}(t_1,t_{\bar{1}})&=&\delta_{\cal C}(t_1-t_{\bar{1}})+\int_{\cal C}\textup{d}\bar{t}\,\vec{\Sigma}(t_1,\bar{t})\,\vec{g}(\bar{t},t_{\bar{1}})\;,\\
\label{KBE2mf}
\vec{g}(t_1,t_{\bar{1}})\left[\,-i\partial_{t_{\bar{1}}}-\vec{h}(t_{\bar{1}})\,\right]&=&\delta_{\cal C}(t_1-t_{\bar{1}})+\int_{\cal C}\textup{d}\bar{t}\,\vec{g}(t_{1},\bar{t})\,\vec{\Sigma}(\bar{t},t_{\bar{1}})\;,
\end{eqnarray}
where the single-carrier energy is given by $h_{mn}(t_1)=\int\textup{d}^dr_1\,\phi^*_m(\vec{r}_1)h(1)\phi_n(\vec{r}_1)$, and in the adjoint equation (\ref{KBE2mf}) the operators are acting to the left. The explicit form of the self-energy matrix $\Sigma_{mn}(t_1,t_{\bar{1}})$ at the HF level is
\begin{eqnarray}\label{HFse}
 \Sigma^\mathrm{HF}_{mn}(t_1,t_{\bar{1}})&=&-i\lambda\,\delta_{\cal C}(t_1-t_{\bar{1}})\sum_{k,l=0}^{n_b-1}[w_{mn,kl}-w_{ml,kn}]\,g^{<}_{kl}(t_1,t_{\bar{1}})\;,
\end{eqnarray}
with the two-electron integrals $w_{mn,kl}$ defined as
\begin{eqnarray}\label{2ei}
w_{mn,kl}^{}&=&\!\int\!\!\!\int \textup{d}^dr\,\textup{d}^d\bar{r}\, \phi^*_m(\vec{r})\,\phi^*_k(\bar{\vec{r}})\,w(\vec{r}-\bar{\vec{r}})\,\phi_n(\vec{r})\,\phi_l(\bar{\vec{r}})\;.
\end{eqnarray}
The detailed structure of $\vec{\Sigma}(t_1,t_{\bar{1}})=\vec{\Sigma}^{\mathrm{HF}}(t_1,t_{\bar{1}})+\vec{\Sigma}^{\mathrm{corr}}(t_1,t_{\bar{1}})$ in second (order) Born approximation is given e.g.~in Refs.~\cite{balzer08,dahlen06,dahlen07}.

For the real-time arguments in $\vec{g}(t_1,t_{\bar{1}})$, it is useful to introduce relative and centre of mass (c.m.) variables, $t_\mathrm{c.m.}=(t_1+t_{\bar{1}})/2$ and $t_\mathrm{rel.}=t_1-t_{\bar{1}}$. The Green's functions with respect to the c.m.~time then account for the statistical (thermodynamic) properties of the artificial atom, see the definitions~(\ref{density}) to (\ref{HFenergy}), while quantities with respect to $t_{\mathrm{rel.}}$ carry the dynamical (spectral) information, cf.~$a(\omega)$ as defined at the end of Sec.~\ref{sec3.2}. Moreover, we note that Eqs.~(\ref{KBE1mf}) and (\ref{KBE2mf}) can, in HF approximation, be further simplified to a single-time equation involving only $t_{\mathrm{c.m.}}$. More technical and numerical details for the time-propagation of the Green's function matrix are to be found in Refs.~\cite{thesis07,dahlen06}. 

\subsection{Dynamical quantities\label{sec3.2}}
The spatial one-particle density in the AA and the HF orbital-resolved occupation probability of state $m$ are
\begin{eqnarray}
\label{density}
\mean{\hat{n}}(\vec{r}_1,t_1)&=&-iG(1,1^+)=-i\sum_{m,n=0}^{n_b-1}\phi^*_m(\vec{r}_1)\,\phi^0_n(\vec{r}_1)\,g_{mn}^<(t_1,t_1)\;,\\
\label{occprob}
\mean{\hat{n}_m}(t_1)&=&g_{mm}^<(t_1,t_1)\;.
\end{eqnarray}
In a conserving approximation, the electron number $N(t_1)=\sum_m \mean{\hat{n}_m}(t_1)$ in the AA is preserved in agreement with the continuity equation, i.e.~$\partial_{t_{1}}\langle\hat{n}\rangle(1)+\textup{div}\langle\vec{j}\rangle(1)=0$ with current density $\langle\vec{j}\rangle(1)=-\frac{1}{2}\{[\nabla_{\vec{r}_1}-\nabla_{\vec{r}_{\bar{1}}}]\,G^<(\vec{r}_1 t_1,\vec{r}_{\bar{1}}t_{\bar{1}})\}_{\vec{r}_1=\vec{r}_{\bar{1}}}$.

The relevant energies involved are the kinetic and potential energy
\begin{eqnarray}
\langle\hat{E}_\mathrm{kin+pot}\rangle(t_1)&=&-i\sum_{m,n=0}^{n_b-1}[t_{mn}+v_{mn}(t_1)]\,g_{nm}^<(t_1,t_1)\;,
\end{eqnarray}
with corresponding definitions $t_{mn}=\int\textup{d}^dr_1\,\phi^*_m(\vec{r}_1)\,[\,-\nabla^2_{\vec{r}_1}/2\,]\,\phi_n(\vec{r}_1)$ and $v_{mn}(t_1)=\int\textup{d}^dr_1\,\phi^*_m(\vec{r}_1)\,[\,\vec{r}_1^2/2\,+\,\gamma(t_1)\vec{r}_1\,]\,\phi_n(\vec{r}_1)$, the HF energy
\begin{eqnarray}
\label{HFenergy}
\langle\hat{E}_\mathrm{HF}\rangle(t_1)&=&-\frac{i}{2}\sum_{m,n=0}^{n_b-1}\Sigma^{\mathrm{HF}}_{mn}(t_1)\,g_{nm}^<(t_1,t_1)\;,
\end{eqnarray}
and the correlation energy $\langle\hat{E}_\mathrm{corr}\rangle(t_1)=-\frac{i}{2}\int_{\cal C}\textup{d}\bar{t}\,\textup{Tr}\{\vec{\Sigma}^{\mathrm{corr}}(t_1,\bar{t})\,\vec{g}(\bar{t},t_1^+)\}$.

Other interesting quantities are e.g.~the time-dependent dipole moment $\langle\hat{d}\rangle(t_1)=-i\,e\int d^2 r_1\{\vec{r}_1\,G^<(\vec{r}_1 t_1,\vec{r}_{\bar{1}} t_1)\}_{\vec{r}_1=\vec{r}_{\bar{1}}}$ or the one-particle spectral function\cite{balzer08} $a(\omega,t_\mathrm{c.m.})=i\sum_m\int dt_\mathrm{rel.}\,e^{i\,\omega\, t_\mathrm{rel.}}[g^>_{mm}(t_1,t_{\bar{1}})-g^<_{mm}(t_1,t_{\bar{1}})]$.

\begin{figure}[t]
\begin{center}
\includegraphics[width=0.625\textwidth]{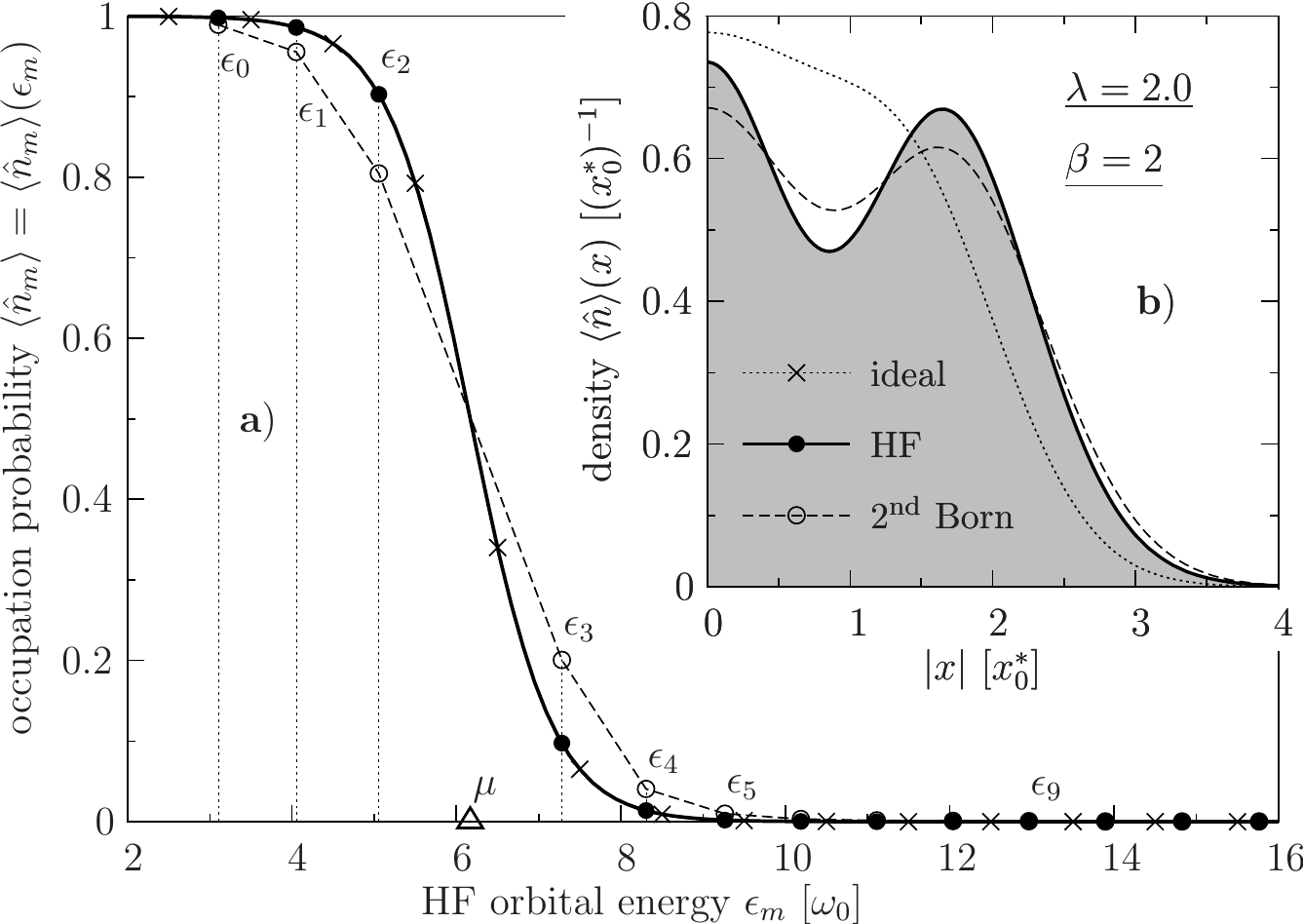}
\end{center}
\caption{Thermodynamic properties of $N=3$ charge carriers in the 1D AA with $\lambda=2.0$ and $\beta=2$. \textbf{a}) HF distribution function $f_\beta(\epsilon_m-\mu)$ (solid curve) and renormalization at the second Born level (dashed curve). The triangle marks the position of the chemical potential in HF approximation. Further, the ideal energies $\epsilon_m^0=m+1/2$ (marked with crosses) for $\lambda\equiv0$ are shifted by the HF value of $\mu$. \textbf{b}) Density profile $\langle\hat{n}\rangle(x)$ for the ideal system as well as for the HF and second Born approximation.}\label{Fig2}
\end{figure}

\section{Numerical results\label{sec4}}
\begin{figure}[t]
\begin{center}
\includegraphics[width=0.95\textwidth]{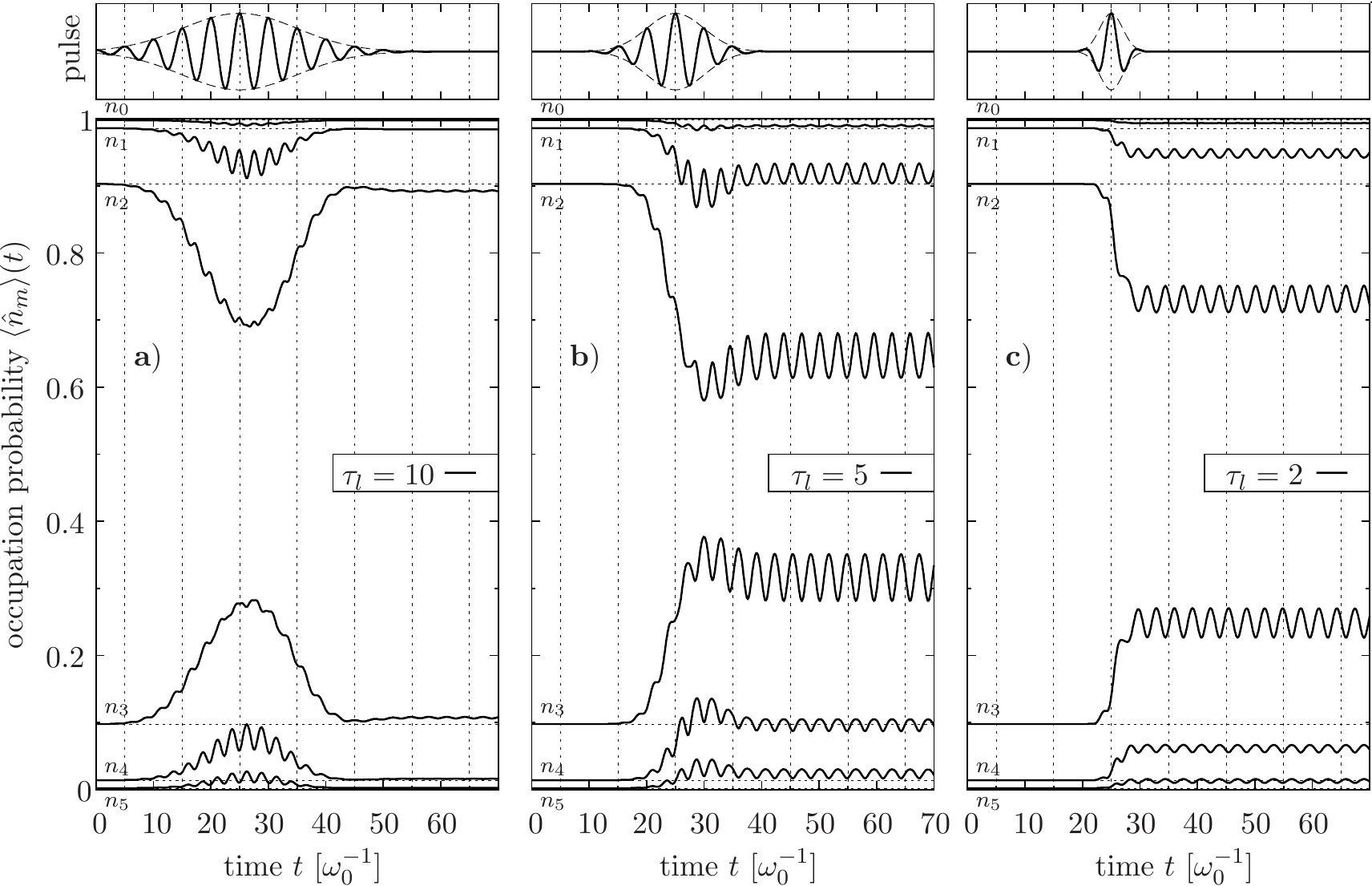}
\end{center}
\caption{Nonequilibrium behavior of the $N=3$ AA system (1D) with parameters $\lambda=2.0$ and $\beta=2$. \textbf{a}) to \textbf{c}) show the mean-field dynamics of the HF orbital occupation probabilities $\mean{n_i}(t)$ at a near-resonant laser frequency $\omega_l=1.25\omega_0$ and three different pulse durations $\tau_l$. The corresponding pulse shapes $\gamma(t)$ are indicated above the figures.}\label{Fig4}
\end{figure}

In this section, we study the dynamical properties of a three-electron AA, when, initially in thermodynamic equilibrium, the system is excited by a single few-cycle laser pulse described in dipole approximation, cf.~Eq.~(\ref{hamx0}). More precisely, the field is linearly polarized in $x$-direction and has the time-dependence
\begin{eqnarray}\label{gamma}
 {\gamma}(t)&=&{\cal E}_0\,e^{-(t-t_l)^2/(2\tau_l^2)}\cos(\omega_l(t-t_l))\;,
\end{eqnarray}
where ${\cal E}_0={\cal E}_l/\sqrt{2 \pi}$ denotes the amplitude of the electric field, the Gaussian envelope is centred at $t_l$, the pulse duration (variance) is given by $\tau_l$, and the oscillation frequency is $\omega_l$, cf.~Figs.~\ref{Fig4} \textbf{a})-\textbf{c}).

\begin{figure}[t]
\begin{center}
\includegraphics[width=0.95\textwidth]{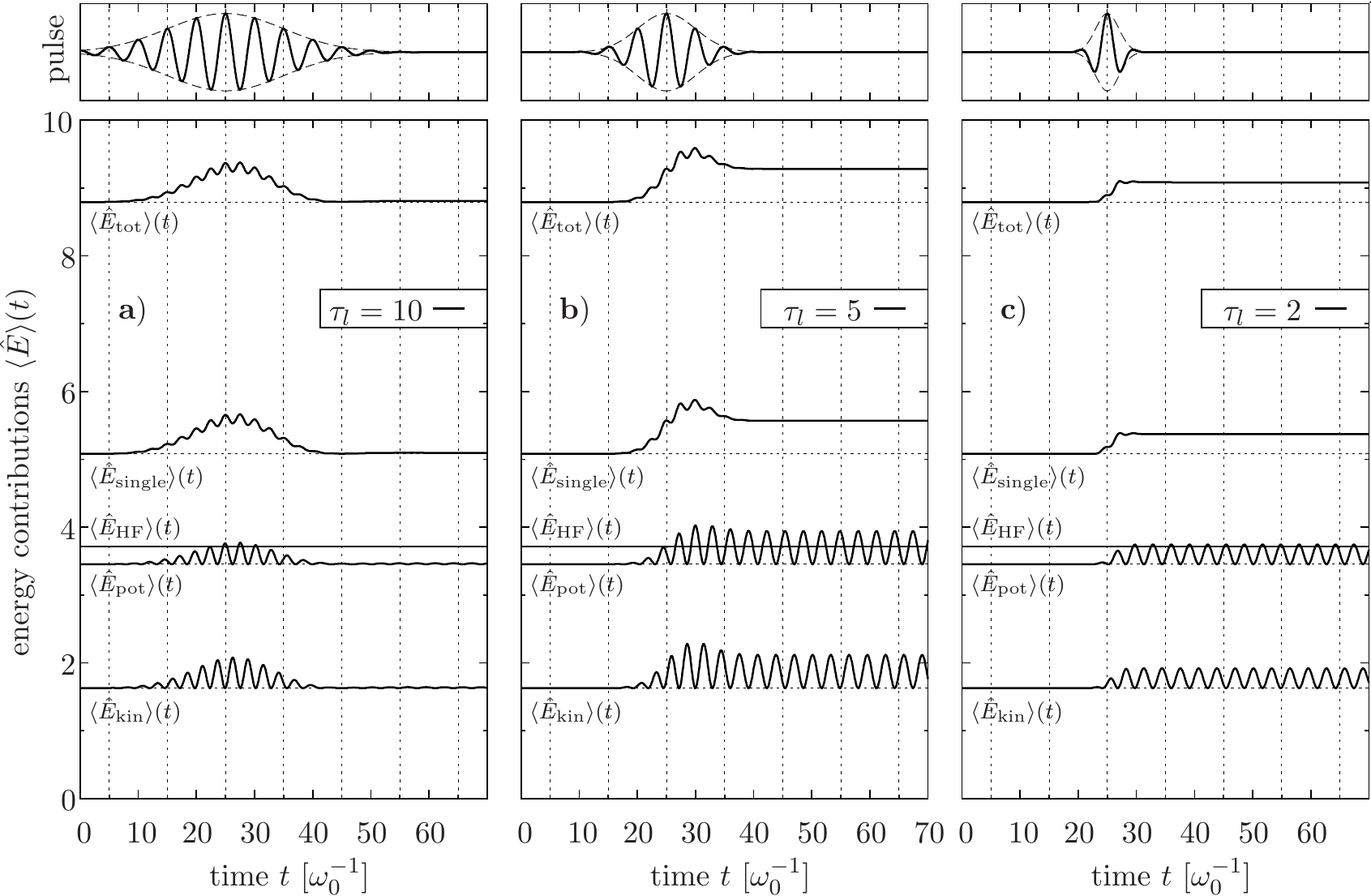}
\end{center}
\caption{For the laser irradiated artificial atom [$\omega_l=1.25\omega_0$] also considered in Fig.~\ref{Fig4}, \textbf{a}) to \textbf{c}) show the mean-field dynamics of the relevant energy contributions in dependence of the pulse duration. For $\tau_l=10$, the AA returns after the excitation close to its initial state (off-resonant situation). Whereas, in Figs.~\textbf{b}) and \textbf{c}) the spectral width of the laser frequency is essentially increased compared to \textbf{a}), leading to the AA remaining in an excited state of the Kohn mode (resonant case). Potential and kinetic energy thereby oscillate out of phase with exactly double confinement frequency $\omega_0$, while the HF energy stays constant.}\label{Fig5}
\end{figure}

\begin{figure}[t]
\begin{center}
\includegraphics[width=0.525\textwidth]{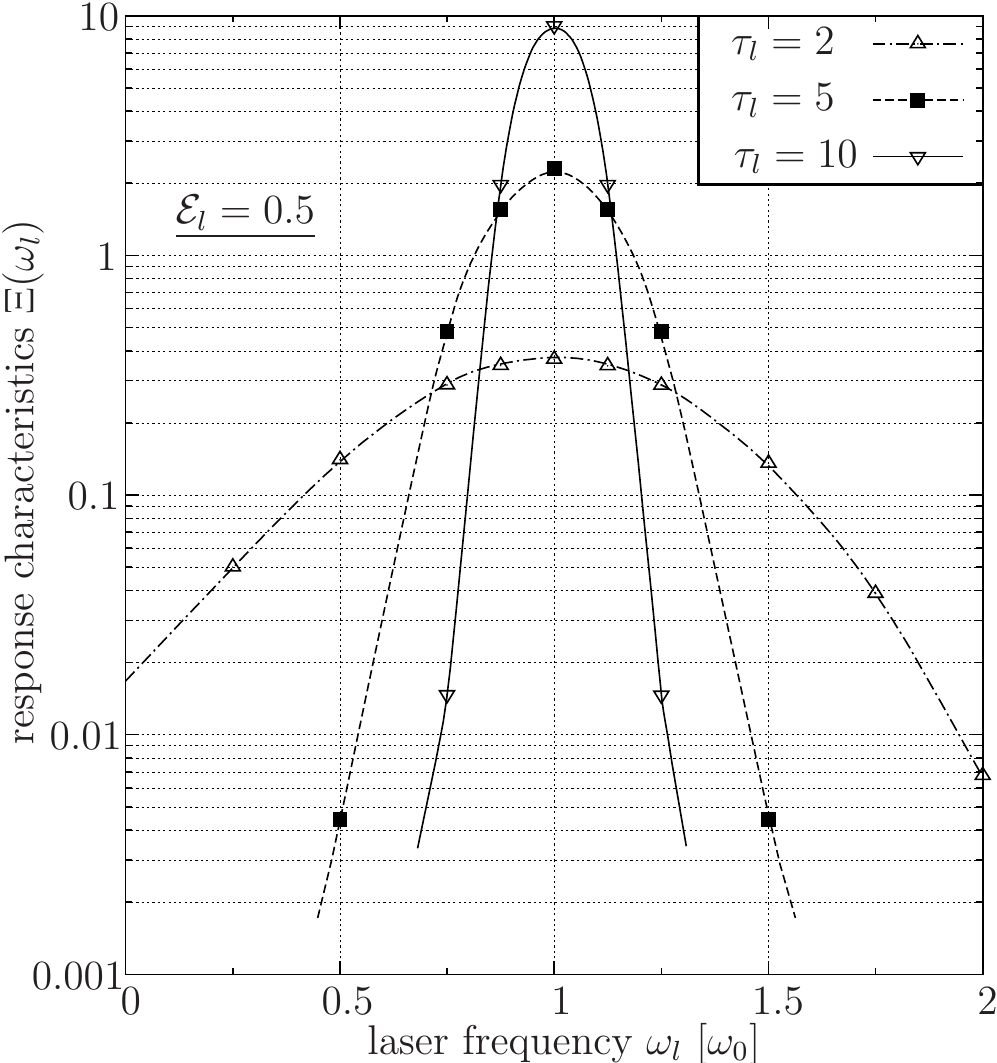}
\end{center}
\caption{Response characteristics $\Xi(\omega_l)$, Eq.~(\ref{defxi}), for the pulsed laser excitation of $N=3$ electrons in the 1D artificial atom. The system parameters are as in Fig.~\ref{Fig2}. The AA shows resonance behavior at the confinement frequency only, i.e.~ for $\omega_l=\omega_0$, and responds via the c.m.~motion (Kohn mode)---rigid translation of the whole density. With increasing pulse durations $\tau_l$ (sharpened laser frequency) the resonance curves become more and more peaked.}\label{Fig3}
\end{figure}

As the response characteristics $\Xi(\omega_l)$ of the quantum system we define the amount of energy that has been absorbed from the laser field for a fixed frequency $\omega_l$, i.e.
\begin{eqnarray}\label{defxi}
\Xi(\omega_l)=\langle\hat{E}_\mathrm{tot}\rangle_{\omega_{l}}(t\rightarrow\infty)-\langle\hat{E}_\mathrm{tot}\rangle(0)\;.
\end{eqnarray}
This quantity together with the time-dependent occupation probabilities $\langle\hat{n}_m\rangle(t)$ allows for the determination of (off)resonant nonequilibrium behavior (and nonlinear effects), see Secs.~\ref{sec4.1} and \ref{sec4.2}.

The nonequilibrium behavior of the quantum system (\ref{ham}) is theoretically well known: Driven by the laser field, the AA exactly responds according to the excitation of the centre of mass (Kohn or sloshing) mode. This is obtained from the Kohn theorem, and its generalization to the case of an additional external dipole field, see Ref.~\cite{bonitz07} and references therein. Its statement is that, independent of dimensionality, the centre of mass coordinate $\vec{R}(t)=N^{-1}\sum_i\vec{r}_i(t)$ of a parabolically confined, interacting electron system performs (equivalently to a single particle in the AA) the motion of a forced harmonic oscillator, $\ddot{|\vec{R}|}+\omega_0^2 |\vec{R}|=Ne E(t)/m_e^*$. Furthermore, this effect is accompanied by a rigid translation of the density profile $\mean{\hat{n}}(\vec{r})$, since the particle interaction appears only in the relative Hamiltonian and $[H_{\mathrm{c.m.}},H_{\mathrm{rel.}}]=0$. The key point in the present study is, however, that the Kohn theorem also holds when the interaction is treated approximately, as long as density, total energy and momentum are preserved (conserving approximation), and also applies to zero and finite temperatures---for the proof see Ref.~\cite{bonitz07}.

The following mean-field results for 1D and 2D have been obtained from NEGF calculations with up to $n_b=40$ HF orbitals. The main limitations of the approach are thereby (i) the basis size, which sets the dimension of the time-evolution matrix $\vec{U}(t_1)=\exp(-i\,[\vec{\vec{h}}(t_1)+\vec{\Sigma}^{\mathrm{HF}}(t_1)]\,t_1)$ to be computed in each time-step (diagonalizing $\vec{h}+\vec{\Sigma}^{\mathrm{HF}}$), and (ii) the two-electron integrals $w_{ij,kl}$, Eq.~(\ref{2ei}), that generally require large memory resources [scaling with $n_b^4$] and need to be processed very frequently in the self-energy expression $\vec{\Sigma}(t_1,t_{\bar{1}})$. With more than $50000$ time-steps needed to achieve convergence, this results in computing times of typically several hours on a single machine. For the correlated time-evolution of the AA in second Born approximation, the propagation must be carried out in the whole two-time plane $(t_1,t_{\bar{1}})$. This is an even more intricate task as one needs to compute all the higher order collision integrals on the r.h.s.~of the KBE. However, these calculations are currently near completion---examples are to be found in Ref.~\cite{thesis07} and for applications on real atoms and small molecules see Refs.~\cite{dahlen05,dahlen06,dahlen07}.



\subsection{1D case\label{sec4.1}}
For the 1D AA calculations in equilibrium and nonequilibrium, we, in Eq.~(\ref{2ei}) have replaced the pure Coulomb interaction
 $w(x-\bar{x})$ by $\lambda\,[(x-\bar{x})^2+\alpha^2]^{-1/2}$ with $\alpha=0.1$ as a regularization parameter. This is necessary to make the integrals $w_{mn,kl}$ finite and, in a physical interpretation, allows for a small transversal spread of the one-electron wave functions\cite{jauregui93}. In 2D, we used $\alpha\equiv0$ as the integrals converge.

We consider three electrons in a 1D artificial atom at $\beta=2$. With $\lambda=2.0$ the system is tuned into the crossover regime between Fermi liquid-like and crystal-like behavior. The equilibrium properties can be read from Fig.~\ref{Fig2} for the HF and second Born approximation. For the HF energy spectrum and the corresponding distribution function $f_\beta(\epsilon_m-\mu)$ of the equilibrium (initial) state, including its collisional renormalization in second Born approximation, see Fig.~\ref{Fig2}~\textbf{a}). The one-electron density $\langle\hat{n}\rangle(x)$ is displayed in Fig.~\ref{Fig2}~\textbf{b}). Compared to the HF result (solid curve, $\langle\hat{E}_\mathrm{tot}^\mathrm{HF}\rangle=8.790$), here, the inclusion of electron-electron scattering (dashed curve, $\langle\hat{E}_\mathrm{tot}^\mathrm{2ndB}\rangle=8.941$), leads to a considerable reduction of the density modulation, which is accompanied by an increase of the total energy of about $1.7$ \%.

Starting from the HF Green's function $G^M(\vec{r}_1,\vec{r}_ {\bar{1}};\tau)$, the AA was now propagated in time under the presence of a laser field (centred at $t_l=25$) with amplitude ${\cal E}_l=0.5$ and frequency $\omega_l=1.25\omega_0$. What happens to the orbital occupations and the energies for different pulse durations $\tau_l$ is shown in Figs.~\ref{Fig4} and \ref{Fig5}. In all cases, gradually, the HF orbitals $m<3$ become depopulated and the states $m\geq3$ analogously become populated with preservation of $N(t)$. Oscillations of the increased total, kinetic and potential energy and $\langle\hat{n}_m\rangle(t)$ thereby occur with twice the confinement frequency. In Figs.~\ref{Fig4}~\textbf{a}) and \ref{Fig5}~\textbf{a}), respectively, the laser excitation is such that the $N$-particle dynamics is decelerated and almost freezed after the pulse has passed. Consequently, we nearly recover the initial state characterized by $\langle\hat{n}_m\rangle(0)$ and $\langle\hat{E}_\mathrm{tot}\rangle(0)$. Also, for different pulse durations, i.e.~different spectral profiles of the laser, the maximum laser energy absorption is observed at the confinement frequency $\omega_0$, cf.~the response function $\Xi(\omega_l)$ in Fig.~\ref{Fig3}. In addition, the single resonance-peak at $\omega_l=\omega_0$ sharpens with the increase of $\tau_l$. Consider now the spatial dynamics. We observe in all cases, that the center of mass of the AA $\vec{R}(t)$ performs a harmonic oscillation with frequency $\omega_0$ while the whole density profile itself is translated rigidly. Accompanying this fact, $E_\mathrm{HF}$ is constant in time, see Fig.~\ref{Fig5}. Thus, we numerically confirm that the Kohn theorem is satisfied.

\begin{figure}[t]
\begin{center}
\includegraphics[width=0.875\textwidth]{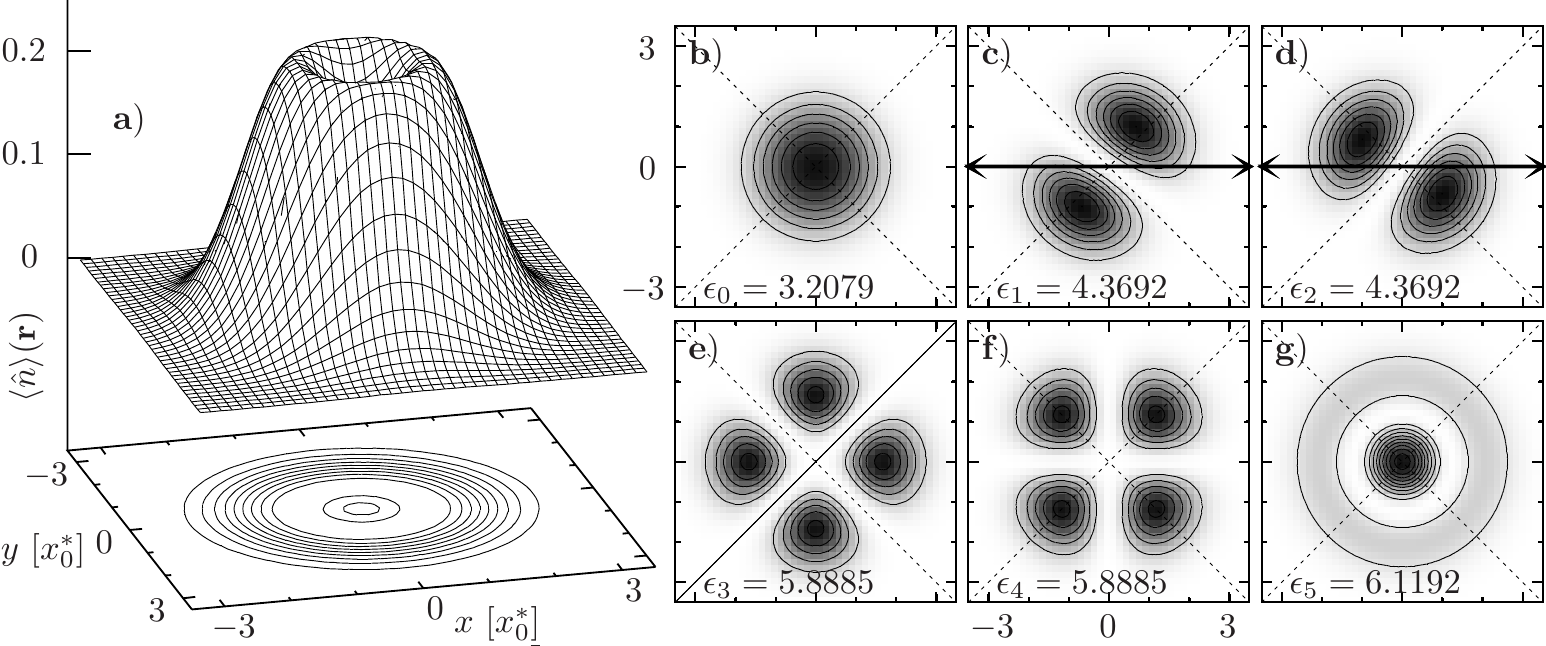}
\end{center}
\caption{Thermodynamic initial state of the 2D artificial atom ($N=3$) at $\lambda=2.0$ and $\beta=2$ (in HF approximation). \textbf{a}) Single-electron density profile $\langle\hat{n}\rangle(\vec{r})$ which is rotationally symmetric. Figs.~\textbf{b}) to \textbf{g}): Energetically lowest spatial, unrestricted HF states $\phi_m(\vec{r})$ with orbital energies $\epsilon_0,\ldots,\epsilon_5$ where $\epsilon_1$ and $\epsilon_2$  as well as $\epsilon_3$ and $\epsilon_4$ are degenerate). The arrows in \textbf{c}) and \textbf{d}) mark the direction of polarization of the laser field $\gamma(t)$, Eq.~(\ref{gamma}).}\label{Fig6}
\end{figure}

\begin{figure}[t]
\begin{center}
\includegraphics[width=0.775\textwidth]{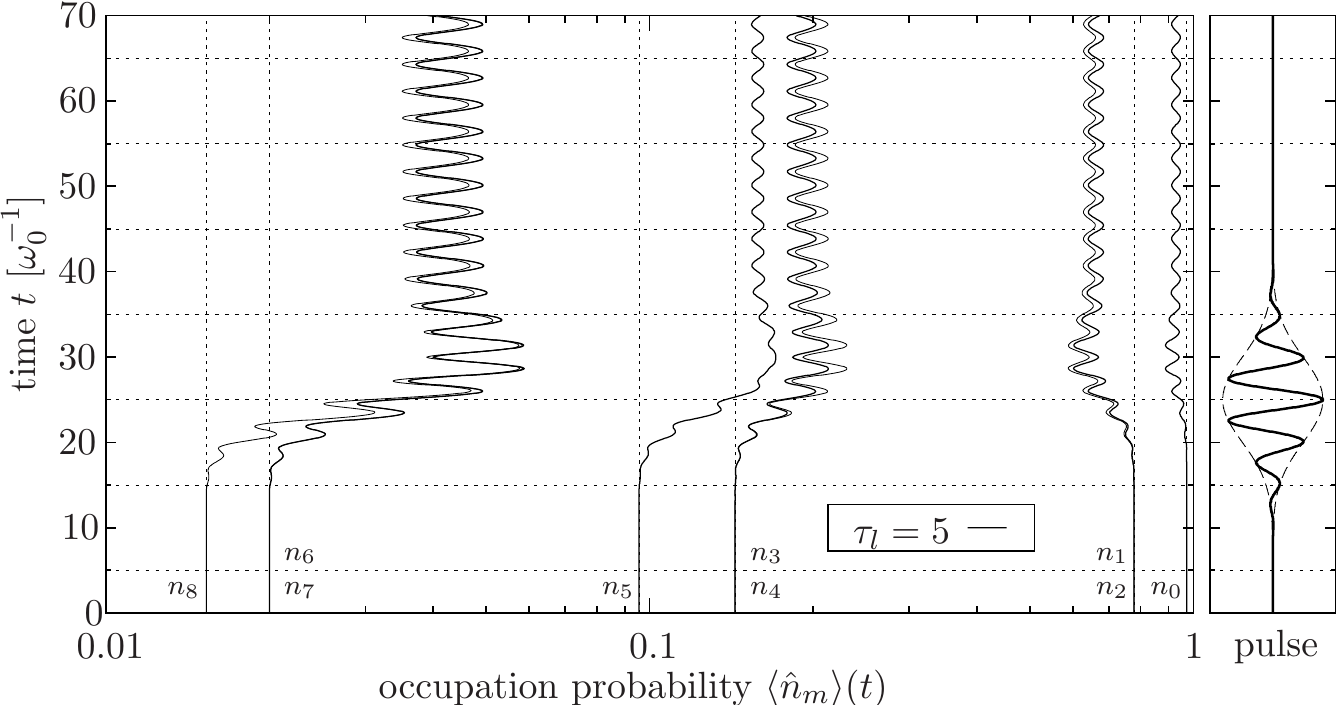}
\end{center}
\caption{Mean-field dynamics of the HF orbital occupation probabilities $\mean{\hat{n}_m}(t)=n_m$ for three charge carriers in a 2D artificial atom with coupling parameter $\lambda=2.0$ and inverse temperature $\beta=2$. The laser frequency is again near-resonant, $\omega_l=1.25\omega_0$, and the pulse duration is $\tau_l=5$. In the initial state the occupation numbers $n_1$ and $n_2$, $n_3$ and $n_4$ as well as $n_6$ and $n_7$ are practically pairwise degenerate, compare with the energy spectrum displayed in Fig.~\ref{Fig6}.}\label{Fig7}
\end{figure}

\subsection{2D case\label{sec4.2}}
For the three-electron AA in 2D, we have chosen the same system parameters, $\lambda=2.0$ and $\beta=2$---however, no regularization parameter $\alpha$ was needed in the two-electron integrals $w_{mn,kl}$. The (almost) rotationally symmetric density profile for the equilibrium state is shown in Fig.~\ref{Fig6} and indicates a ring-like structure. It is instructive to note, that the unrestricted HF solution of the Dyson equation leads to orbitals $\phi_m(\vec{r})$, that are in general arbitrarily oriented in space, compare Fig.~\ref{Fig6} \textbf{b}) to \textbf{g}). Together with the energetically degenerate states $m=1$ and $2$ ($m=3$ and $4$, etc.), this has the following consequence on the dipole excitation: As degenerate orbitals can be differently oriented relative to the laser field, in the time-evolution of the artificial atom this degeneracy is lifted. In the present case, the orbitals $m=1$ and $m=2$ (see Fig.~\ref{Fig6} \textbf{c}) and \textbf{d})) are almost aligned with the diagonals (dotted lines), nevertheless the small deviations are sufficient to clearly influence the evolution of the occupation numbers $\langle\hat{n}_m\rangle(t)$, see $\langle\hat{n}_1\rangle$ and $\langle\hat{n}_2\rangle$ as well as $\langle\hat{n}_3\rangle$ and $\langle\hat{n}_4\rangle$ in Fig.~\ref{Fig7}.



\section{Conclusion and outlook}
We have presented an analysis of femtosecond relaxation of few-particle quantum dots during and after a short laser pulse. The method of NEGF wa shown to be efficient to describe the dynamics even in the range of strong Coulomb correlations. Numerically, the C.m.~mode excitation can serve as a very sensitive test for the NEGF calculation [and any other numerical code] involving quantum many-body approximations Ref.~\cite{bonitz07}.

\section*{Acknowledgements}
We acknowledge stimulating discussions with R.~van~Leeuwen, A.~Filinov and S.~Bauch. This work was supported by the Innovationsfonds Schleswig-Holstein.

\end{document}